\documentclass[11pt,a4paper]{article}
\usepackage[truedimen,margin=32mm]{geometry} 

\usepackage{titlesec}
\titleformat*{\section}{\large\bfseries}
\titleformat*{\subsection}{\it}

\usepackage{mathrsfs}
\usepackage{booktabs}
\usepackage{amssymb} 
\usepackage{amsmath}
\usepackage{mathtools}
\usepackage{ascmac}
\usepackage{amsthm}
\usepackage[pdftex]{color}
\usepackage{natbib}
\usepackage{setspace}
\usepackage{bm}
\usepackage{url}
\usepackage{color}
\usepackage{tabularx}
\usepackage{ulem}

\usepackage{tikz}
\usetikzlibrary{cd}

\usepackage{algorithm}
\usepackage{algorithmic}

\usepackage[hang,small,bf]{caption}
\usepackage[subrefformat=parens]{subcaption}
\theoremstyle{definition}

\newcommand{\mG}{\mathcal{G}}

\newcommand{\mP}{\mathcal{P}}

\newcommand{\mN}{\mathcal{N}}

\renewcommand{\hat}{\widehat}
\renewcommand{\tilde}{\widetilde}

\usepackage{color}

\def\diag{\mathop{\rm diag}\nolimits}

\title{\bf Spatiotemporal factor models for functional data with application to population map forecast}
\author{}
\date{}

\begin{document}

\maketitle
\doublespacing

\vspace{-1.5cm}
\begin{center}
{\large 
Tomoya Wakayama$^{\ast}$\footnote{Corresponding author, Email: tom-9@g.ecc.u-tokyo.ac.jp}
and Shonosuke Sugasawa$^{\dagger}$
}

$^{\ast}$Graduate School of Economics, The University of Tokyo\\
$^{\dagger}$Faculty of Economics, Keio University
\end{center}

\begin{abstract}
The proliferation of mobile devices has led to the collection of large amounts of population data.
This situation has prompted the need to utilize this rich, multidimensional data in practical applications. In response to this trend, we have integrated functional data analysis (FDA) and factor analysis to address the challenge of predicting hourly population changes across various districts in Tokyo. 
Specifically, by assuming a Gaussian process, we avoided the large covariance matrix parameters of the multivariate normal distribution.
In addition, the data were both time and spatially dependent between districts.
To capture these characteristics, a Bayesian factor model was introduced, which modeled the time series of a small number of common factors and expressed the spatial structure through factor loading matrices.
Furthermore, the factor loading matrices were made identifiable and sparse to ensure the interpretability of the model. 
We also proposed a Bayesian shrinkage method as a systematic approach for factor selection. Through numerical experiments and data analysis, we investigated the predictive accuracy and interpretability of our proposed method. We concluded that the flexibility of the method allows for the incorporation of additional time series features, thereby improving its accuracy.

\end{abstract}

\noindent%

\section{Introduction} \label{sec:intro}

With the proliferation of mobile devices, an increasing amount of population data is being collected, and there is a growing demand for its use.
Currently, we can quickly determine the amount of people staying in Tokyo, Japan, at any given time or place.
These population data can be utilized to reduce crowding and traffic congestion through transportation planning, improve the efficiency of rideshares and delivery services, promote consumption, and guide evacuation and estimate casualty losses during disasters \citep{wang2018spatial,suzuki2013using,paez2004spatial}.

The population data collected by NTT Docomo, one of the largest mobile carriers in Japan, is especially interesting.
NTT Docomo has approximately 82 million customers (excluding corporate accounts) in Japan, and based on their operational data, the number of mobile terminals in each base station area is counted. The population of each area is then extrapolated with high accuracy using NTT Docomo's cell phone penetration rate \citep[See][for more details]{terada2013population,oyabu2013evaluating}. We will focus on the five special wards of Tokyo as our study area. A mesh is defined as a square of 500 meters, and there are approximately 400 meshes in the area. For each mesh, hourly population data was obtained for 365 days.

\begin{figure}[t]
  \centering
  \includegraphics[width=9cm]{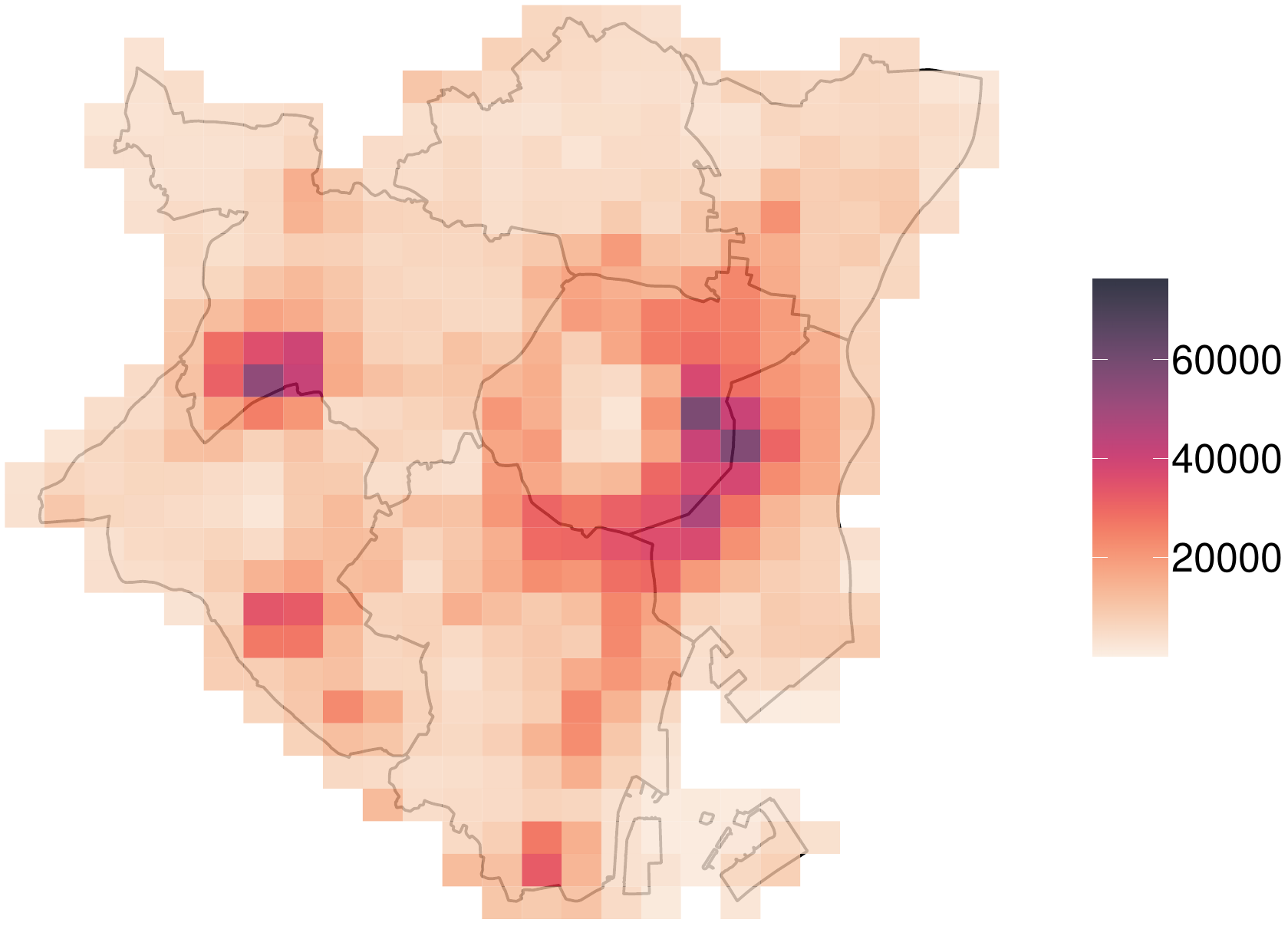}
  \caption{Number of people in central districts of Tokyo at 14:00 on January $29, 2019.$}\label{fig:districts}
\end{figure}

\begin{figure}[t]
  \centering
  \includegraphics[width=12cm]{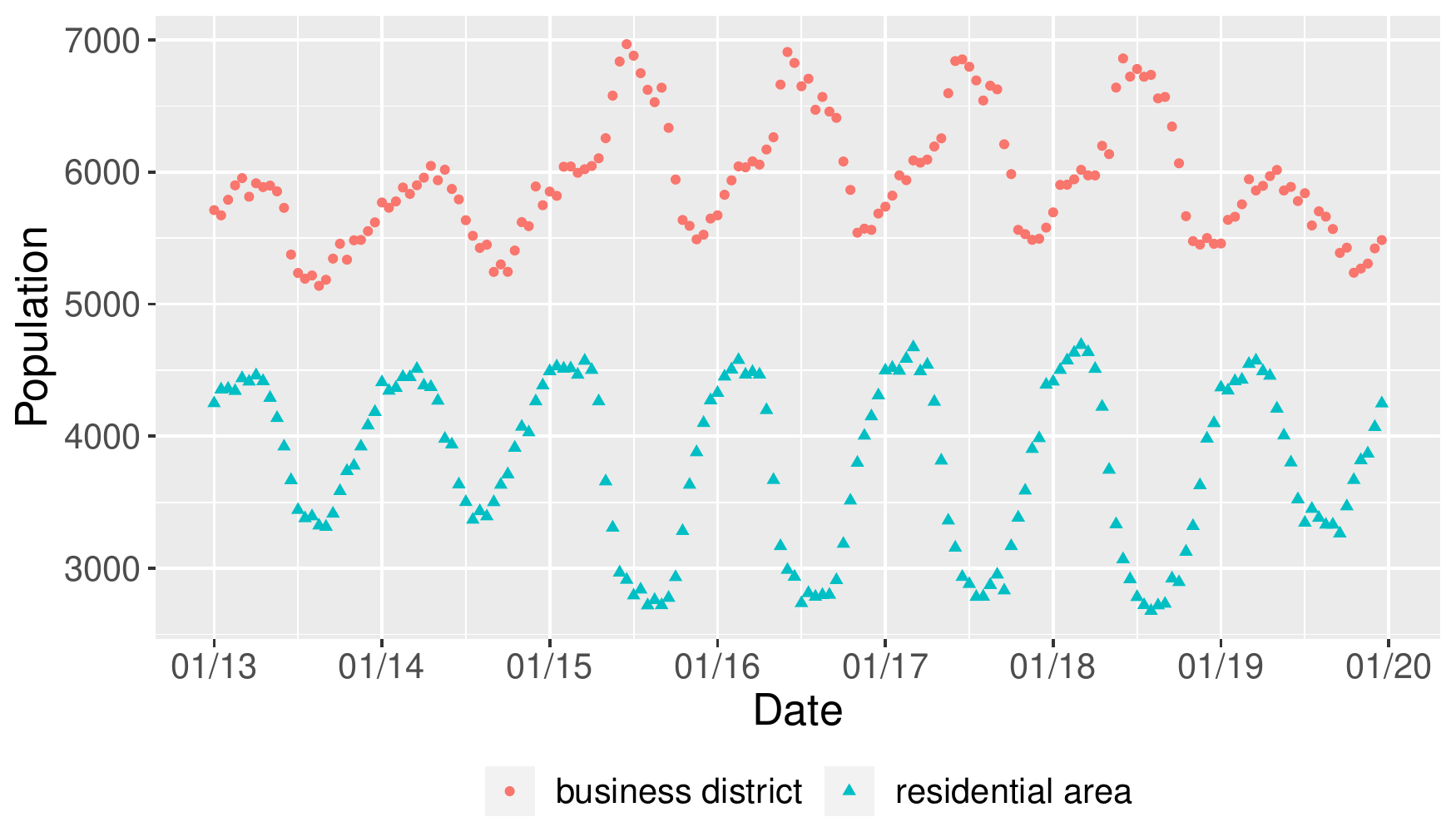}
  \caption{Hourly population data in two districts. Red represents a business district; blue represents a residential area.}\label{fig:population}
\end{figure}

Our objective in this study is to predict the population of each district. We introduce key characteristics of the data that must be understood before constructing the model; the first is the spatial structure.
Figure \ref{fig:districts} illustrates the number of people at 14:00 on January 29, 2019, in each district of Tokyo. There are some districts with more people and some with fewer people, and these geographic changes are gradational. Hence the spatial correlation should be taken into account.
Another critical feature is the time series structure.
Consider the hourly population transition in two districts, an office and a residential area, for the week beginning Sunday, January 13, 2019, as shown in Figure \ref{fig:population}. The red and blue points represent the flow of people in a business district and residential area, respectively.
Basically, the population trends of the previous day are the same as those of the following day, but the population trend switches drastically between holidays (Sunday, Saturday, and Monday, which is a public holiday) and weekdays.
This is intuitive; on weekdays, more people stay in the business area, whereas on holidays, downtown and residential areas are relatively more populated.
In addition, the data presented here has the distinction of being large-scale, with dozens-dimension data collected over hundreds of days across numerous locations. This is a considerable obstacle in spatiotemporal modeling.

To resolve those issues, we consider a novel integration of (a) functional data analysis (FDA), and (b) Bayesian factor models.
\begin{itemize}
    \item[(a)] FDA is a methodology that treats and analyzes longitudinal data as curves, reduces parameters, and facilitates the handling of high-dimensional data \citep{ramsay2004functional,horvath2012inference,kokoszka2017introduction}. Even with discretely measured data, it is natural to think of the data as if there is a latent curve because the data are assumed to exist not only at the point of observation but also at other points.
    By assuming the path of the Gaussian process to be the underlying function, we reduce the number of parameters and analyze them.
    \item[(b)] To efficiently estimate the mean parameters of the Gaussian process, we introduced the Bayesian factor model \citep[e.g.,][]{calder2007dynamic,nakajima2013bayesian,lopes2000bayesian,lopes2003expected}. Based on the state space model, a few distinctive districts of the city are assigned as factors to reduce the computational cost because only the time series of factors needed to be considered. In addition, the factors described the temporal structure through state evolutions and the factor loading matrix captures the spatial correlation structure among districts.
\end{itemize}
These two elaborations make it possible to implement the large-scale spatiotemporal model. The method is feasible by a Gibbs sampler \citep{gelfand1990sampling} and also allows for the development of factor selection schemes based on posterior predictive loss (PPL) \citep{gelfand1998model}.
Furthermore, the factor loading matrices are set to be identifiable, and are estimated to be sparse. Identifiability, achieved by using a Cholesky-type matrix, yields unique inference results for factor loading matrices. Sparsity, made possible by newly incorporating a shrinkage prior distribution, allows us to identify which districts influence other districts.
This interpretability, along with the uncertainty inherent in Bayesian models, makes predictions important in applications.
Highly explanatory forecasts are effective for convincing decision-makers, and information such as 95\% probability of bad case scenarios is of value to them.
Since the demand for their use will continue to increase as more detailed regional and temporal population data become available, our proposed method may contribute to evidence-based policy making.

Several previous studies have addressed the FDA framework for spatio-temporal data \citep{zhang2023bayesian,li2021multilevel,wakayama2021locally,romano2011clustering,giraldo2011ordinary,jiang2012clustering}. However, most of these have taken the approach of using time as the argument of the function and incorporating it into the analysis of spatial function data, ignoring the temporal structure.
While there have been a few spatiotemporal developments in topics unrelated to forecasting (e.g., missing value completion by \cite{zhu2022spatiotemporal}), our contribution to the field, spatiotemporal FDA, is to propose a forecasting model that reflects spatiotemporal features in the state space. In addition, many univariate analyses of spatio-temporal data have been studied \citep{banerjee2014hierarchical,prado2021time}. Also, numerous spatio-temporal methods have also been studied for univariate data. Our method can be regarded as a generalization of these ideas to functional responses with some innovations (e.g., factor loading design), and therefore has wide applicability beyond its current use.

The remainder of the paper is organized as follows. 
Section \ref{sec:model} describes the settings, model, and factor selection procedure.
In Section \ref{sec:sim}, we study the features and performance of our method compared with other methods through numerical experiments.
We apply our method to population flow data in Section \ref{sec:DA}. The contributions of this study are discussed in section \ref{sec:con}.
{ 
Appendix provides the Bayesian computation procedure.
}

\section{Spatiotemporal factor models for functional data}\label{sec:model}
\subsection{Setting and model}\label{sec:set}
Let $y_{ts}(\tau)$ be the observed functional data (population) at time $t\in \{1,\ldots,T\}$ and in region (mesh, in our application,) $s\in\{1,\ldots,N\}$ with a measurement point $\tau \in \{\tau_1,\ldots,\tau_K\}$ of function. { In the example presented in Section~\ref{sec:intro}, $t$ and $\tau_k$ represent a day and a time within the day, respectively, hence defining a curve with $K$ measurements (time within the day).} For any $t$ and $s$, we assume the following measurement error model.
\begin{align*}
    y_{ts}(\tau) &=  z_{ts}(\tau)+ \varepsilon_{ts}  , \ \ \ \ \  \varepsilon_{ts}\sim N(0,e_s^2),
\end{align*}
where { $\{\varepsilon_{ts}\}_{t,s}$ are error terms that are all independent of each other}, $e^2_s$ is an unknown variance, and $z_{ts}$ is the focus. 
These models are widely adopted in the context of Bayesian modeling of functional data \citep{yang2016smoothing,yang2017efficient,jiang2012clustering,wakayama2022functional}.
Assume function $z_{ts}$ follows the Gaussian process.
\begin{align*}
    z_{ts}(\tau) \sim \mathcal{GP}(f_{ts}, \eta_s^2 R(\phi_s)),
\end{align*}where $f_{ts}$ is the mean parameter, $R(\phi_s)= \rho_{\phi_s}(d)$ is correlation kernel and $\eta_s$ is its scale.
Given the observed points, $\tau_1,\ldots,\tau_K$, the above assumption leads to the following multivariate normal distribution: 
\begin{align*}
    (z_{ts}(\tau_1),\ldots, z_{ts}(\tau_K)) \sim N \left((f_{ts}(\tau_1),\ldots, f_{ts}(\tau_K)), \tilde{R}_s \right),
\end{align*}
where $\tilde{R}_s$ denotes a $K\times K$ Gram matrix with $(i,j)$-components $\eta_s^2 \rho_{\phi_s}(|\tau_i-\tau_j|)$.

Viewing a vector as a finite subset of a stochastic process is beneficial. If we attempt to estimate a $K$-dimensional covariance matrix using ordinary multivariate analysis, as many as $K \times (K-1)/2$ (e.g., $24(24-1)/2=276$ if $K=24$) parameters are required, which is laborious to estimate.
However, by assuming that the vector is a finite subset of the path of a stochastic process, we only need to estimate a few parameters of the covariance kernel (only $\eta_s$ and $\phi_s$ for each point in the above case).
That is, time-consuming calculations are eliminated by considering the underlying stochastic process.

\subsection{State space factor models}
To model the mean parameters over time and space, the following state models are considered:
\begin{align}\label{eq:factor}
     \bm{z}_t &= (B\otimes I_K)\bm{x}_t+\bm{\nu}_t,\ \ \ \ \ \  \bm{\nu}_t\sim N(\bm{0},\mathrm{blockdiag}(\tilde{R}_1,\ldots, \tilde{R}_N) ) \\
    \bm{x}_{t}  &= G \bm{x}_{t-1}+ D_t\bm{\mu} + \bm{\omega}_t , \ \ \ \  \bm{\omega}_t\sim N(\bm{0},\Lambda)\nonumber,
\end{align}where $\bm{z}_{t}:=\left(\bm{z}_{t1},\bm{z}_{t2},\ldots,\bm{z}_{tN} \right) :=\left(z_{t1}(\tau_1),\ldots,z_{t1}(\tau_K),z_{t2}(\tau_1),\ldots,z_{tN}(\tau_K) \right)$ is an $NK$-dimensional vector,
$\bm{x}_{t}:=\left(\bm{x}_{t1},\bm{x}_{t2},\ldots,\bm{x}_{tM} \right) :=\left(x_{t1}(\tau_1),\ldots,x_{t1}(\tau_K),x_{t2}(\tau_1),\ldots,x_{tM}(\tau_K) \right)$ is an $MK$-dimensional vector, $\bm{\mu}:=\left(\bm{\mu}_{1},\bm{\mu}_{2},\ldots,\bm{\mu}_{M} \right) :=\left(\mu_{1}(\tau_1),\ldots,\mu_{M}(\tau_K) \right)$ is an $MK$-dimensional vector,
$D_t\in\{-1,0,1\}$ is a dummy variable,
$G$ is an $MK\times MK$ state evolution matrix,
$\bm{\nu}_t$ and $\bm{\omega}_t$ are the corresponding error terms, 
$\Lambda:=\mathrm{diag}(\lambda^2_1\ldots,\lambda_M^2)$ is error variance in factor time series and { $B$ is an $N\times M$-matrix with a hierarchical structural constraint~\citep{prado2021time, Aguilar1999a} of the form}
\begin{align*}
    B &=\left(
        \begin{array}{ccccc}
          1 & 0& 0&\ldots  &0\\
          b_{21} & 1 & 0&\ldots  &0\\
          b_{31} & b_{32} & 1&\ldots  &0\\
          \vdots & \vdots &  &\ddots  &\vdots\\
          b_{M1} & b_{M2} & b_{M3}&\ldots  &1\\
          b_{M+1,1} & b_{M+1,2} & b_{M+1,3}&\ldots  &b_{M+1,M}\\
          \vdots & \vdots &  &\ddots  &\vdots\\
          b_{N,1} & b_{N,2} & b_{N,3}&\ldots  &b_{N,M}\\
        \end{array}
        \right).
\end{align*}
Note that the number of factors $M$ is less than $N$.
This formulation is a factor model \citep{Aguilar1999a, elkhouly2021dynamic, gamerman2008spatial}.
At each $t$, a large number of vectors $\{\bm{z}_{ts}\}_{s=1}^N$ is represented by a small number of vectors $\{\bm{x}_{ts}\}_{s=1}^M$.
This makes it compatible with large-scale data because only a small number of evolutions must be considered.

Furthermore, defining the factor loading matrix in this way guarantees identifiability and interpretability.
Analysis might be easier if all components were parameters, but there could be multiple expressions describing the relationship between explanatory factors and explained variables, which would render the parameters inexplicable.
In this case, for example, the first factor $\bm{x}_{t1}$ is equal to the first mean parameter $\bm{z}_{t1}$ minus noise; the second factor $\bm{x}_{t2}$ is equal to the second mean parameter $\bm{z}_{t2}$ minus $b_{21}\times \bm{x}_{t1}$ minus noise, and so on.
That is, the $s$th factor represents the part of the $s$th trend that is not explained by the first $s-1$th factors. { 
In other words, if the other regions are similar to the first, they take a high value in the first column; if they are not so similar (have different characteristics from the first), they take a higher value in the subsequent columns.}
Hence, we formulated this method to clarify the role of each parameter.

Note that the time series equation contains the $D_t\bm{\mu}$ term, where $D_t$ depends on a combination of $t$ and $t-1$.
$D_t$ is $1$ when $(t,t-1)$ is (holiday, weekday), $-1$ for (weekday, holiday), and $0$ otherwise.
$\mu_s$ represents the difference in holidays compared with weekdays at each location, which allows for the modeling of day-off effects. This term can be designed more flexibly based on empirical knowledge. For example, the day before a holiday, such as Friday, tends to have a different population trend from that on a typical weekday \citep{stutz2004charting,lu2012strategic}; hence, a term can be added to reflect this:
\begin{align}\label{eq:add}
     \bm{z}_t &= (B\otimes I_K)\bm{x}_t+\bm{\nu}_t,\ \ \ \ \ \  \bm{\nu}_t\sim N(\bm{0},\mathrm{blockdiag}(\tilde{R}_1,\ldots, \tilde{R}_N) ), \\
    \bm{x}_{t} &= G \bm{x}_{t-1}+ D_t\bm{\mu} + D_t' \bm{\mu}'+ \bm{\omega}_t , \ \ \ \ \bm{\omega}_t\sim N(\bm{0},\Lambda)\nonumber,
\end{align}
where $D_t'$ and $\bm{\mu}'$ are analogous of $D_t'$ and $\bm{\mu}'$, respectively.
The benefits of this flexibility are discussed in Section~\ref{sec:DA}.

\subsection{Factor loading matrix}
To reflect the spatial structure, we consider the column vector $\bm{b}_{\cdot s}:= (b_{s+1,s},\ldots,b_{N,s})$ of $B$ and set the following prior:
\begin{align}\label{def:col}
    \bm{b}_{\cdot s}|\upsilon\theta_s  \sim \mN(\bm{0},\upsilon\theta_s^2 Q_s^{-1}(\psi)),\ \ \ \ \ \ s=1,\ldots,M
\end{align}where $\upsilon$ and $\theta_s$ are scale parameters, $Q_s(\psi)= (I-\psi W_s)(I-\psi W_s)^{\top}$, $\psi$ is a spatial autoregression parameter and $W_s$ is the adjacency matrix for $s+1$ to $N$th points, whose $(i,j)$-entry is one if district $i$ and district $j$ are adjacent and zero otherwise. This formulation is known as the CAR model \citep{banerjee2003hierarchical}, which is designed such that adjacent districts are similarly affected by a factor in this context. The $\psi$ represents their similarity. For example, if $\psi$ is $0$, then elements of $\bm{b}_{\cdot s}$ are independent and, consequently, exhibit a low similarity. 

{
Another strategy for introducing the prior on $B$ is studied by \cite{SHIN2023100763}. Despite the difference in the type of data handled - univariate versus functional data - this work and our approach share the same modeling philosophy. In both methods, a prior distribution is imposed on the columns of the factor loading matrix in order to introduce spatial dependence. But we note that \cite{SHIN2023100763} imposes the Gaussian prior on $(b_{1,s},\ldots,b_{N,s})$ rather than the random elements of the column vector. This allows considering the dependency between fixed values $(b_{1,s},\ldots,b_{s,s})$ and random variables $(b_{s+1,s},\ldots,b_{N,s})$. However, the calculation of the posterior distribution needs the "sampling under a hard constraint" approach, increasing the computational cost. In contrast, to effectively address the large-scale dataset, we assume the prior distribution merely on $(b_{s+1,s},\ldots,b_{N,s})$.
}

The covariance matrix of $\bm{b}_{\cdot s}$ (the effect of the $s$th factor) relies on $Q_s$.
In other words, adjacent districts tend to be similarly affected.
Additionally, designing $Q_s$ as a sparse matrix makes computation less expensive because fast inverse matrix computation techniques as well as fast random sampling from a multivariate normal distribution are developed (e.g., ``sparseMVN" package in R language).

To facilitate the interpretability of the dependencies between districts and factors, we adopt the following prior distribution as the scale parameters: \begin{align*}
    \upsilon \sim C^+(0,1), \ \ \ \ \theta_s\sim C^+(0,1), \ \ \ s=1,...,M,
\end{align*}
where $C^+$ denotes the half-Cauchy prior. 
Such prior is used in the horseshoe prior \citep{carvalho2009handling,carvalho2010horseshoe} for a univariate parameter, and the resulting distribution of $\bm{b}_{\cdot s}$ is a multivariate version of the horseshoe prior.
A similar multivariate prior is adopted in \cite{shin2020functional} and \cite{wakayama2022functional} in non-spatial settings. 
The horseshoe distribution is known for its strong shrinking ability, which allows the coefficients of singular factors to be zero.
In addition, unlike the Laplace distribution, it has a property called tail robustness, which firmly leaves the non-shrinking parts large.
This clarifies whether the factor is effective and also allows for highlighting important venues or specific facilities.

\subsection{Selecting factors}\label{subsec:select}
The critical concern in this section is how to determine the factors. Even in univariate problems, determining the factors and number of factors is difficult. { The typical method is to align those that appear important based on domain knowledge \citep{prado2021time}. However, this approach requires a subjective selection of factors, and it is unclear whether the selected factors are crucial. First, we select several factors in arbitrary order according to domain knowledge and what the analyst wishes to know. The appropriate factors are then chosen using the criteria and a shrinkage prior, as follows.
}
{ 
\begin{itemize}
    \item[1.] Prepare some sets of factors as candidates. 
    \item[2.] Assign shrinkage priors to $\upsilon$ and $\theta_s$ and implement the proposed method (for smaller scale data) for all candidates.
    \item[3.] Calculate the PPLs and choose the best set from the candidates.
\begin{align*}
{\rm PPL} = \sum_{s=1}^N \Biggl[\sum_{t=1}^T \left\{ \bm{y}_{ts}-{\rm E}_p[\bm{z}_{ts}]\right\}^\top
\left\{ \bm{y}_{ts}-{\rm E}_p[\bm{z}_{ts}]\right\}+TK{\rm E}_p[\sigma_s^2] + \sum_{t=1}^T\mathrm{tr} \left({\rm Cov}_p(\bm{z}_{ts})\right) \Biggr]
\end{align*}
    \item[4.] In the chosen set, if the coefficient of the $s$th factor $\bm{b}_{\cdot s}$ is unshrunk, we consider the $s$th factor necessary but delete it otherwise.
\end{itemize}
}

Because $\bm{b}_{\cdot s}$ is the coefficient corresponding to the $s$th factor, the $s$th factor does not affect the others (i.e., it is insignificant) if it is reduced to zero by the shrinkage distribution.
Thus, we assign more factors to the model beforehand, and retain the essential factors and eliminate unnecessary factors using the effect of the shrinkage distribution.

Although this procedure allows for factor selection, implementing it on a large-scale dataset is time-consuming. This negates one advantage of the factor model, which is that it reduces the computational burden.
Therefore, we recommend implementing factor selection for smaller-scale data and then implementing the proposed model with the selected factors for the entire dataset.

\section{Numerical experiment}\label{sec:sim}
To confirm the usefulness of the proposed method, we investigated the properties of the proposed method and compared its accuracy with that of existing methods.

\subsection{Experimental setting}
\begin{figure}[tb]
    \centering
\begin{tikzcd}
C_1 \ar[r, "2/3"] 
& C_3 \ar[dr, ""] 
&
&[1.5em] \\
&
&\ \ \ \ \ \  C_6,\ldots,C_{N} \\
C_2 \ar[r, "2/3"] 
& C_4 \ar[ur, ""] &  & C_5\\
\end{tikzcd}
    \caption{Diagram showing the influence of each district on the other districts.}
    \label{fig:diag}
\end{figure}
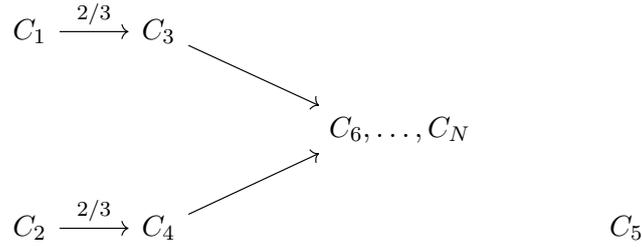

First, we consider the structure of the city. We assume $N$ districts, $C_1 - C_{N}$, where districts $C_1$ and $C_2$ are dominant, influencing the other districts, as shown in Figure \ref{fig:diag}. Assume that $C_3$ and $C_4$ are the next most influential districts and that they have the trends of $C_1$ and $C_2$, but also have their own trends, which spread to adjacent districts $C_6 - C_{N}$. Note that it is unrealistic for a single district, either $C_3$ or $C_4$, to be adjacent to $N-4 \ (\gg10)$ districts; however, we ignore this for the numerical experiment.

The data-generating process is defined as follows.
{ 
\begin{itemize}
\item[-]
Measurement error
$$
y_{ts}(\tau)\sim N(z_{ts}(\tau),e_s^2)
$$

\item[-]
State variables from factors
\begin{align*}
&\ \ \ \ \ \  z_{ts}(\tau) = x_{ts}(\tau) + \mG\mP(0,R(1)/4 ),\ \ \ s=1,2,5 \\
&\ \ \ \ \ \  z_{t3} = 2/3 x_{t1}+ x_{t3}+ \mG\mP(0,R(1)/4), \ \ \ \ \ \  
z_{t4} = 2/3 x_{t2}+ x_{t4}+ \mG\mP(0,R(1)/4), \\
&\ \ \ \ \ \  z_{ts} = w_{1s} z_{t3}+ w_{2s} z_{t4}+ \mG\mP(0,R(1)/4), \ \ \ s=6,...,N,
\end{align*}

\item[-]
Time series of factors
$$
x_{ts}(\tau) = 0.8 x_{t-1,s}(\tau) + N(\bm{0},I_K) , \ \ \ s=1,...,5,\ \ \ \ \ \  x_{1,1},\ldots,x_{1,5}\overset{\mathrm{i.i.d.}}{\sim} \mG\mP(0,25R(4)).
$$
\end{itemize}
}
Here $R(\phi)$ is the radial basis function kernel defined as $R(\phi) = \exp(-\|\tau_i-\tau_j\|^2/\phi)$ and $e_s$ is the measurement error deviation. { $z_{ts}\ (s=6,...,N)$} is the mixture of $z_{t3}$ and $z_{t4}$, and their weights -- $\bm{w}_1=\{w_{1s}\}_{s=6}^N$ and $\bm{w}_2=\{w_{2s}\}_{s=6}^N$ -- are independently sampled from an $(N-5)$-variate Gaussian distribution with mean zero and a band-type covariance matrix, in which the diagonal is $1$, the $(i,i+1)$th and $(i,i-1)$th entries are $1/2$, and the other entries are $0$. The day-off effect is fixed at zero and is not discussed here, as it is the focus of the following section.

All data were measured at $K(=24)$ points on the function over a period of $T$ days. The first $M(=5)$ districts were also considered factors. 
To perform the experiment on different space-time scales, we prepared $(20,50)$ and $(50,90)$ as combinations of $(N,T)$.
We set the noise such that the signal-to-noise ratio (SNR) was the same for all points. That is, the noise variance was heterogeneous. Specifically, we set $e_s$ to $1/5$ (high SNR) and $1/2$ (low SNR) as the standard deviation of the signal in each district.

We conduct the following three methods\begin{itemize}
    \item[-] FFM: our proposed functional factor model.
    \item[-] NSFFM: non-sparse version of the functional factor model. The prior distribution of $B$ is constructed as
\begin{align}
    \bm{b}_{\cdot s}|\theta  \sim \mN(\bm{0},\theta^2 Q_s^{-1}(\psi)),\ \ \ \ \ \ 
    \theta\sim IG(0.1,0.1), \ \ \ \ \ \ s=1,\ldots,M. \nonumber
\end{align} This allowed us to investigate how the sparsity of factor loading matrix $B$ affects interpretability and estimation accuracy.
    \item[-] BART: Bayesian additive regression trees developed by \cite{chipman2010bart}. The purpose of using BART is to study the accuracy of FFM estimation compared to nonparametric flexible methods; however, it is not a time-series method. BART can be applied by ignoring the spatial structure and considering a bivariate regression problem at each location (in this case, the explanatory variables were $t$ and $\tau$).
    \item[-] { UDLM: univariate dynamic linear model~\citep{prado2021time}, a widely used model in time-series forecasting, was used as a competitor. $y_t = \mu_t + v_t;$ $\mu_t = \mu_{t-1} + w_t;$ $v_t \sim N(0, V);$ $w_t \sim N(0, W)$. The UDLM models were implemented using ``Rstan". The prior distributions for $\mu_0$ were standard normal distributions, and those for the error variances $V$ and $W$ were $IG(\frac12,\frac12)$.
    }
\end{itemize}

For all methods, we used $5000$ posterior draws after discarding $15000$ burn-in samples. Because the Gelman and Rubin statistics \citep{Gelman_Rubin_1992} for the parameters were calculated for the four chains and were less than $1.1$, the MCMC was judged to be convergent \citep{Gelman_Hill_2006}. Also, we note that the average effective sample size of the chains is 1409.1.
The resulting sample medians were considered point estimates. To assess the point estimates and sampled posterior distribution, we adopted the following criteria.
\begin{itemize}
\item[-]
Root-mean-square error (RMSE): Difference between the posterior medians and the true values, defined as  
$$
\mathrm{RMSE}=\sqrt{\frac{1}{NTK} \sum_{s=1}^{N}\sum_{t=1}^T\sum_{\tau=1}^{K}  \bigl(\hat{z}_{ts}(\tau)-z_{ts}(\tau)\bigl)^2}.
$$
\item[-] 
Coverage probability (CP): Coverage accuracy of the credible interval, defined as 
$$
\mathrm{CP}=\frac{1}{NTK} \sum_{s=1}^{N}\sum_{t=1}^T\sum_{\tau=1}^{K}  \mathbb{I}_{\{ \hat{z}_{ts}^{97.5}(\tau)>z_{ts}(\tau) >\hat{z}_{ts}^{2.5}(\tau) \}}.
$$
\end{itemize}

\subsection{Result}
Table~\ref{tab:full} lists the RMSEs and CPs for each scenario. The FFM and NSFFM performed better than BART and UDLM.
This is because BART is a nonparametric method that does not consider spatial or time-series structures, whereas FFM is a spatiotemporal method. Hence, the RMSE is lower and CP is higher for FFM than for BART.
{ Also, the generated data is complex to capture by UDLM, as it contains a mixture of periodicity, such as daily periodicity and days of the week, as well as spatial structure.} In addition, under all scenarios, FFM performed slightly better than NSFFM because the spatial structure was captured more accurately by completely eliminating unnecessary factors.

\begin{table}[h]
\caption{Averaged values of root-mean-square error (RMSE) and coverage probability (CP) of 95$\%$ credible interval for functional factor model (FFM), non-sparse functional factor model (FSFFM), Bayesian additive regression trees (BART), and univariate dynamic linear model (UDLM).}
\label{tab:full}
\centering
\medskip
\begin{tabular}{cccccccccccc} 
\toprule
SNR & $(N,T)$ &Method &  & RMSE & CP(\%) \\
\midrule
low & (20,50) &{\bf FFM} & & 0.459 & 96.6 \\  
& &NSFFM &  & 0.465 & 96.8 \\ 
& &BART & & 1.647 & 66.6 \\ 
& &UDLM & & 1.535 & 91.2 \\ 
 & (50,90) &{\bf FFM} & & 0.321 & 97.8 \\  
& &NSFFM & & 0.352 & 98.5 \\ 
& &BART & & 1.346 & 58.6 \\ 
& &UDLM & & 1.601 & 88.3 \\ 
\midrule
high & (20,50) &{\bf FFM} & & 0.192 & 95.5\\  
& &NSFFM & & 0.197 & 97.3 \\ 
& &BART & & 1.341 & 63.3 \\ 
& &UDLM & & 1.128 & 93.2 \\ 
 & (50,90) &{\bf FFM} & & 0.135 & 97.8 \\  
& &NSFFM & & 0.149 & 98.8 \\ 
& &BART & & 1.194 & 55.5 \\ 
& &UDLM & & 1.092 & 93.0 \\ 
\bottomrule
\end{tabular}
\end{table}

\begin{figure}[]
\centering
\begin{minipage}[b]{0.32\linewidth}
    \centering
    \includegraphics[keepaspectratio, height=8cm]{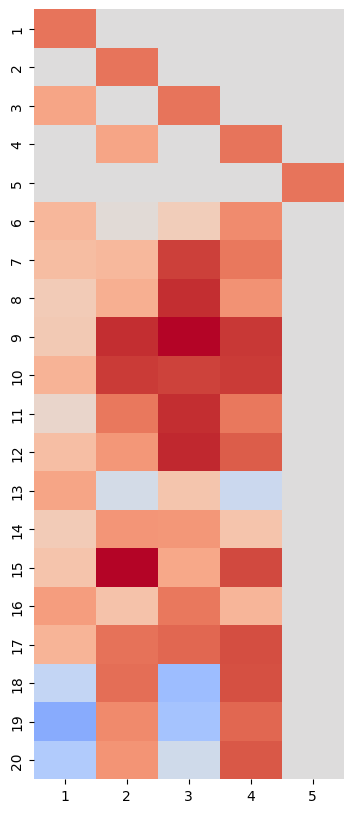}
    \subcaption{True}
  \end{minipage}
\begin{minipage}[b]{0.32\linewidth}
    \centering
    \includegraphics[keepaspectratio, height=8cm]{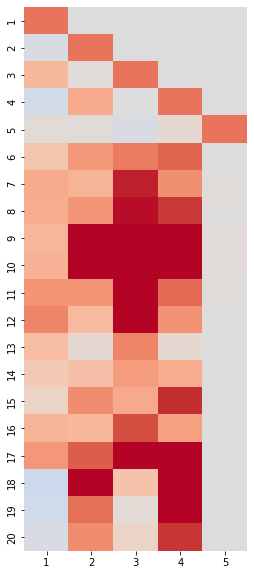}
    \subcaption{FFM}
\end{minipage}
\begin{minipage}[b]{0.32\linewidth}
    \centering
    \includegraphics[keepaspectratio, height=8cm]{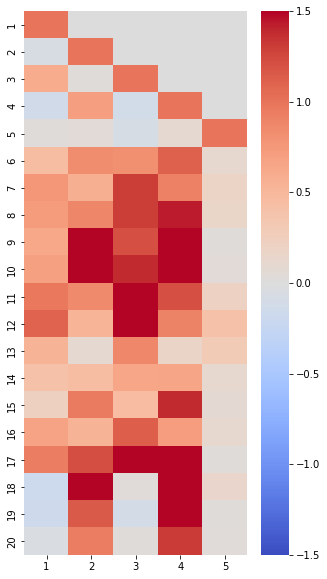}
    \subcaption{NSFFM}
\end{minipage}
   \caption{True factor loading matrices (left), estimated factor loading matrices by functional factor model (middle) and non-sparse functional factor model (right) when signal-to-noise ratio is low.}\label{fig:B}
\end{figure}

\begin{figure}[]
    \centering
  \begin{center}
  \includegraphics[width=12cm]{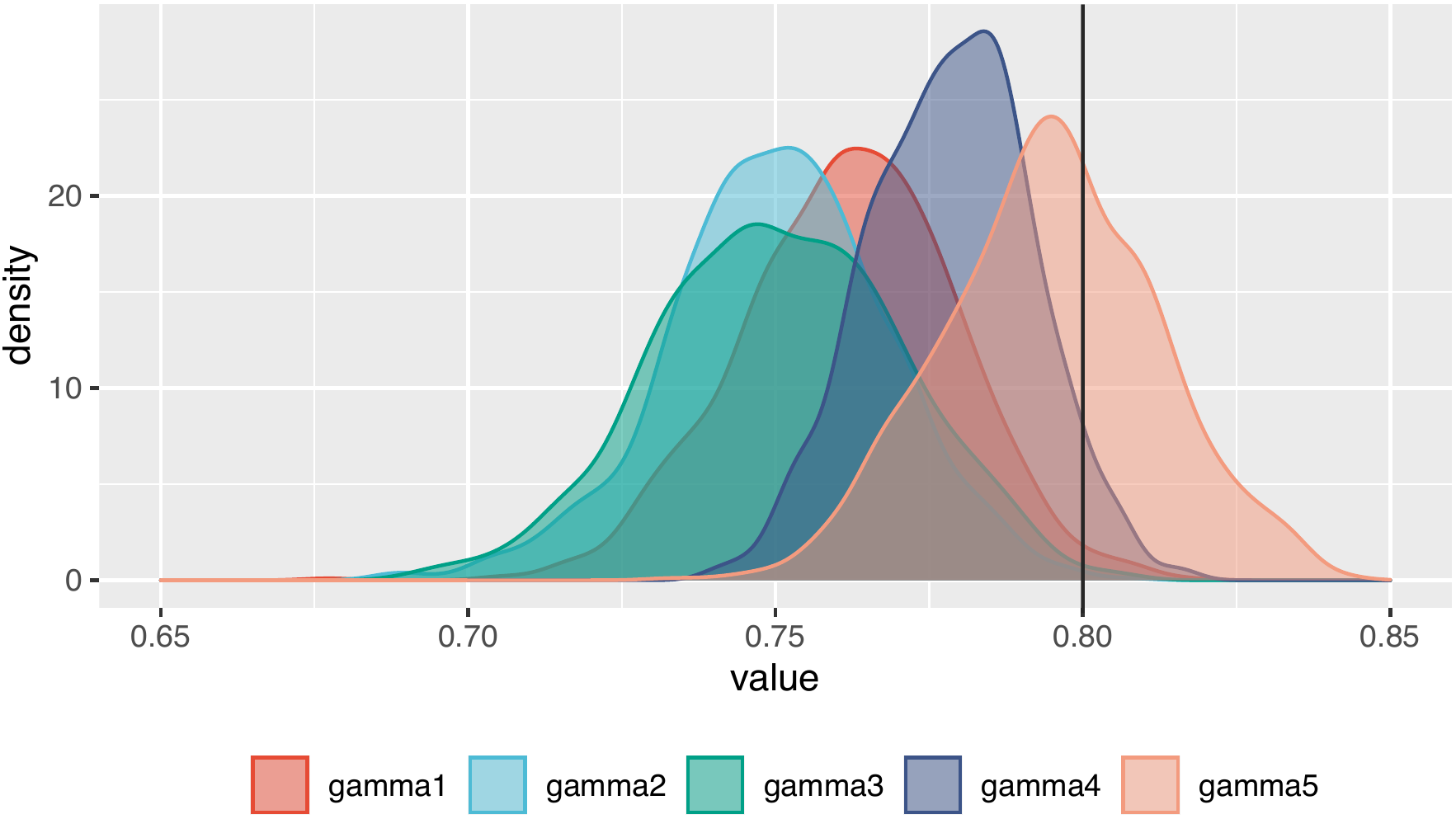}
  \end{center}
  \caption{Posterior distribution of the autoregressive parameters.}
  \label{fig:gamma}
\end{figure}

Next, we discuss the impacts of these factors on each district. Figure \ref{fig:B} shows the estimated factor loading matrices, which represent the influence of five factors on twenty districts. The left plot is estimated using the FFM and the right plot is estimated using the NSFFM. The major difference between the two plots is the coefficient of the fifth factor, that is, the fifth column of matrix $B$. 
The NSFFM results suggest that the fifth factor has a small impact on districts 12 and 13. This is inconsistent with the original data-generating process and is misleading; because NSFFM lacks the ability to sparsify irrelevant factors, it must place some weight on these factors.
By contrast, FFM removes the weights of unnecessary factors. This allowed us to clarify the connection between factors and districts.

We also considered the time-series structure. Figure \ref{fig:gamma} shows the posterior distributions of the autoregressive parameters when $(N,T)$ is $(20,50)$ and SNR is low; the results are similar for the other scenarios. All results were distributed around the true values; therefore, the time-series structure was well modeled. 
In particular, $\gamma_3$ and $\gamma_4$ are close to $0.8$ although they are considered challenging to estimate accurately because, unlike other factors, the 3rd and 4th factors are observed as a mixture of other factors.

\section{Analysis of population data in Tokyo}\label{sec:DA}
This section describes the implementation of the proposed method using real data. 
Because our main aim is accurate prediction, the factors needed to be selected beforehand, the process of which is described in Section \ref{subsec:fac}. Then, before the forecast, Section \ref{subsec:off} discusses the day-off effect, which is important to understand the city.
Next, the forecasting performance and possible improvements are investigated in Section \ref{subsec:pred}.
Because the data were observed hourly and daily for one year, $K=24$ and $T=365$. Saturdays, Sundays, and national holidays in 2019 were defined as days off.
Because different scales at each location would make it challenging to interpret the factor loading matrix, the scale data were normalized as follows: \begin{align*}
    \frac{y_{ts}(\tau)}{\sqrt{\sum_{t,\tau}y_{ts}^2(\tau)/K/T}}.
\end{align*}

\subsection{Factor selection}\label{subsec:fac}
\begin{figure}[t]
\centering
  \includegraphics[width=7cm]{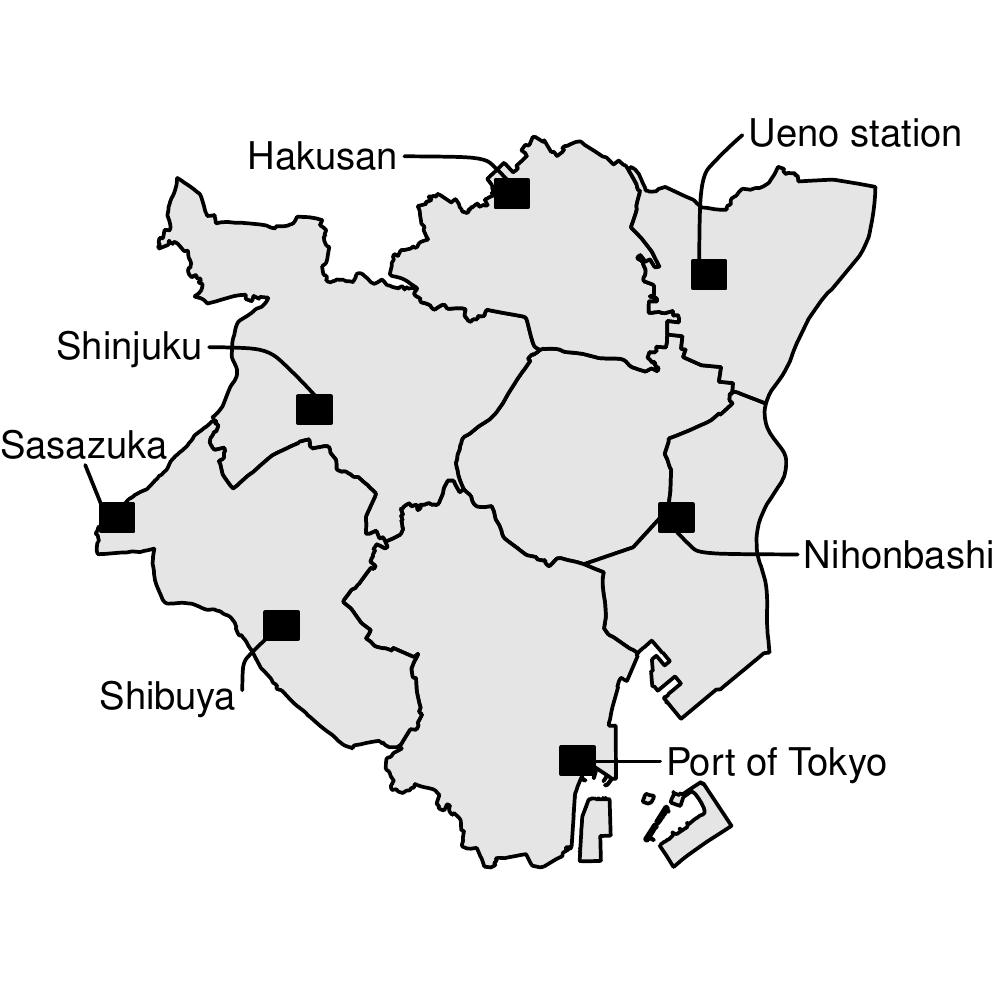}
   \hspace{-0.3cm}
  \caption{Locations of factor districts.}
  \label{fig:facmap}
\end{figure}

The dataset for factor selection comprised population data for the first $100$ days and $50$ randomly selected locations. {
\begin{itemize}
    \item[1.] First, we prepared four sets of factors, each consisting of seven elements. One set was chosen subjectively and three were chosen randomly. 
    \item[2.] For each set, we implemented the proposed methods with $\upsilon \sim C^+(0,1), \ \theta_s\sim C^+(0,1)$, and $s=1,...,M.$
    \item[3.] After calculating the PPL, we chose the set such that the PPL was minimized. This set is listed in Table \ref{description} and illustrated in Figure \ref{fig:facmap}.
\end{itemize}
}

\begin{table}[b]
 \caption{Description of the factor districts and the explained districts.}
 \label{description}
 \centering
  \begin{tabular}{cclll}
  \toprule
   &Number & District & Description  \\
   \midrule
   &1 & Nihonbashi & Business district \\
   &2 & Shibuya  & Downtown \\
   &3 & Sasazuka & Residential area \\
   &4 & Ueno Station & Hub station \\
   &5 & Shinjuku & Downtown \\
   &6 & Hakusan & Residential area \\
   &7 & Port of Tokyo & Seaport \\
  \bottomrule
  \end{tabular}
\end{table}

\begin{figure}[p]
  \centering 
  \includegraphics[height=22cm]{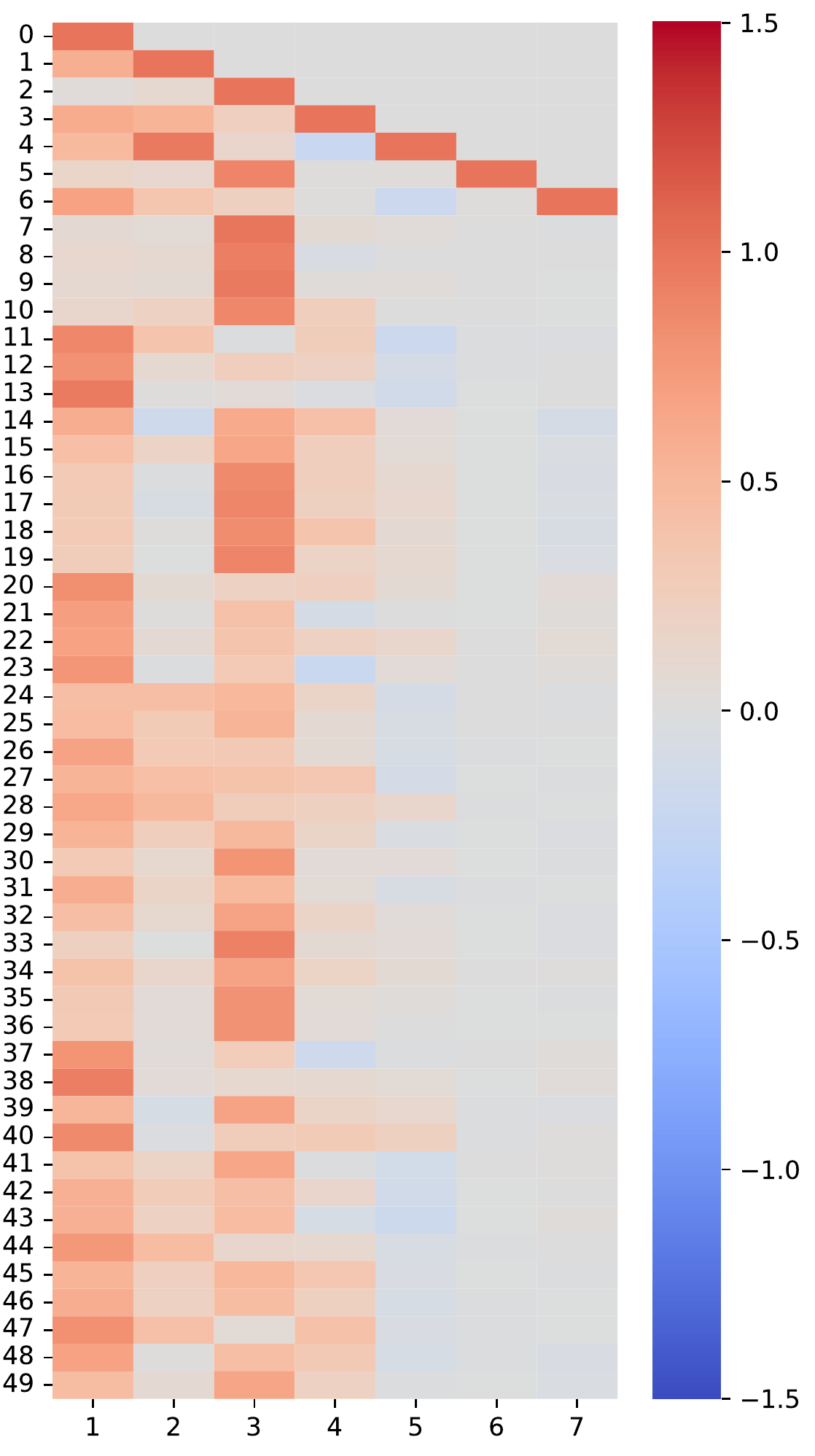}
  \hspace{-1cm}
  \caption{Estimated factor loading matrix by the proposed method.}
  \label{fig:sele}
\end{figure}
Figure \ref{fig:sele} shows the estimated factor loading matrix for the selected set. Several important things can be inferred from this. One is the relationship between the factors and other districts. For example, the $8-11$th and $18-20$th districts are categorized as being residential areas. In Table \ref{description}, the third factor (district) is in a residential area as well. Hence, it is natural for those districts to be well explained by the third factor. Additionally, the $12-15$th districts were office areas, similar to the first district, and the factor loading matrix was consistent with this.
These relationships are intuitively plausible and allow for a more detailed understanding of the urban structure.
Another noteworthy point is shrinkage. Although we selected the best set from the candidates in the above procedure, we could also choose factors within the set. Figure \ref{fig:sele} exhibits that the coefficients of the sixth and seventh factors are negligible.
The seventh factor was based on population data from a wharf in Tokyo, but in this case, there were no districts that could be explained by this factor. The sixth district was in a residential area which matched the third district. Hence, the trends specific to residential areas were captured by the third factor, and there was little that could be additionally explained by the sixth factor.

\begin{figure}[t]
  \begin{center}
  \includegraphics[width=14cm]{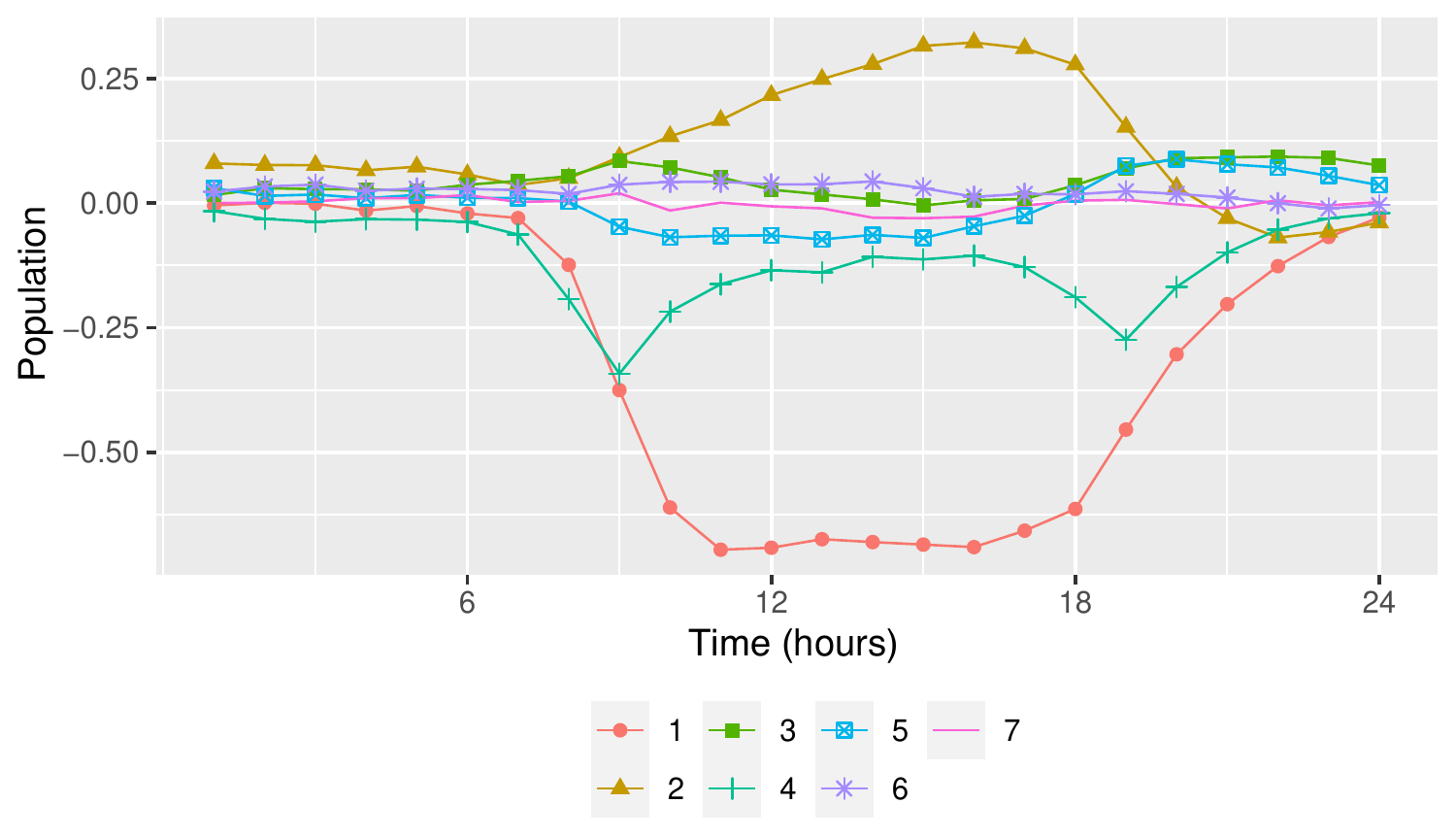}
  \end{center}
  \caption{Day-off effects $\bm{\mu}_s$ of all factors.}
  \label{fig:off}
\end{figure}

\subsection{Day-off effect}\label{subsec:off}
Next, we focused on the day-off effect.
Figure \ref{fig:off} illustrates the estimated day-off effect $\bm{\mu}_s$ of each factor.
These results provide interesting insights.
The first is the day-off effect in the office districts. As represented by the first factor (the business district), there were fewer people on holidays because they did not come to work.
In downtown and residential areas, the holiday population tended to increase.
This is because many people went downtown to enjoy the holidays but did not leave their homes in residential areas to commute to school or work, as they did on weekdays.
Another interesting feature was the day-off effect of the fourth factor. This region was a hub station and many people used it to commute to work or school. Hence, Ueno Station was particularly populated during rush hour (9:00 and 19:00) on weekdays.
However, this did not occur on holidays, so that the population decreased at these times.

\subsection{Prediction}\label{subsec:pred}
Finally, we predicted future data for $400$ districts.
We implemented three experiments. In the first (second, third) one, we used the first $152$ ($212$, $272$) days as training data and the following $30$ days as test data to evaluate the performance. Note that the results reported below are the average of these three experiments.
Based on the discussion in Section \ref{subsec:fac}, we employed five districts in the best set as factors. 
To optimize performance, we considered the following two extensions to the original method:
\begin{itemize}
    \item[I.] Add pre-day-off effects to the factor evolution equation (\ref{eq:add}).
    \item[II.] Add pre-day-off effects and pre-working effects to the factor evolution equation. The latter effect was added because on weekdays before a day off there were likely to be different trends and vice versa.
\end{itemize}
{ As a competitor, we performed the UDLM and chose standard norma prior distributions for the state parameters and $IG(\frac12,\frac12)$ for the error variances.}

To evaluate the prediction error, we define the scale-adjusted RMSE (SRMSE) as follows:
\begin{align*}
    \frac{ \sqrt{T^{-1}\sum_t\|\hat{\bm{y}}_{ts} - \bm{y}_{ts}\|_2^2} }{ \sqrt{T^{-1}\sum_t\|\bm{y}_{ts}\|_2^2} }.
\end{align*}

We applied the proposed methods to train the data and obtained $5000$ posterior samples after $15000$ burn-in periods. For these samples, we obtained forecasts using the Monte Carlo approximation and reported the SRMSE values and $95\%$ coverage probability of $95\%$ prediction intervals.

\begin{table}[tb]
\caption{
Scale-adjusted root mean square error (SRMSE) and Coverage probability (\%) of three methods for each factor, average SRMSE for non-factors, and average SRMSE for all districts on weekdays, holidays, and the entire period.}
\label{tab:RMSE} 
\centering
\medskip
\begin{tabular}{cccccccccccc} 
\toprule
&Factor & Original FFM & Extension I & Extension II & UDLM \\
\midrule
&1 & 0.050 & 0.053 & 0.045 & 0.643\\  
&2 & 0.258 & 0.161 & 0.134 & 0.439\\ 
All days&3 & 0.041 & 0.037 & 0.028 & 0.134\\ 
&4 & 0.112 & 0.097 & 0.087 & 0.402\\  
&5 & 0.306 & 0.206 & 0.190 & 0.295\\ 
&Average& 0.088 & 0.066 & 0.062 & 0.315\\ 
&CP & 94.7 & 97.2 & 97.0 & 87.1\\
\midrule
&1 & 0.039 & 0.041 & 0.041 & 0.870\\  
&2 & 0.200 & 0.089 & 0.096 & 0.458\\ 
Working days&3 & 0.038 & 0.050 & 0.027 & 0.172\\ 
&4 & 0.110 & 0.089 & 0.070 & 0.393\\  
&5 & 0.200 & 0.144 & 0.127 & 0.281\\ 
&Average& 0.079 & 0.058 & 0.058 & 0.233\\ 
&CP & 95.8 & 97.6 & 96.7 & 88.9\\
\midrule
&1 & 0.202 & 0.045 & 0.041 & 0.540\\  
&2 & 0.352 & 0.145 & 0.075 & 0.411\\ 
Days off&3 & 0.044 & 0.044 & 0.033 & 0.151\\ 
&4 & 0.120 & 0.093 & 0.091 & 0.204\\  
&5 & 0.435 & 0.198 & 0.128 & 0.356\\ 
&Average& 0.134 & 0.062 & 0.056 & 0.342\\ 
&CP & 93.5 & 96.9 & 97.1 & 83.6\\
\bottomrule
\end{tabular}
\end{table}

\begin{figure}[t]
  \begin{center}
  \includegraphics[width=14cm]{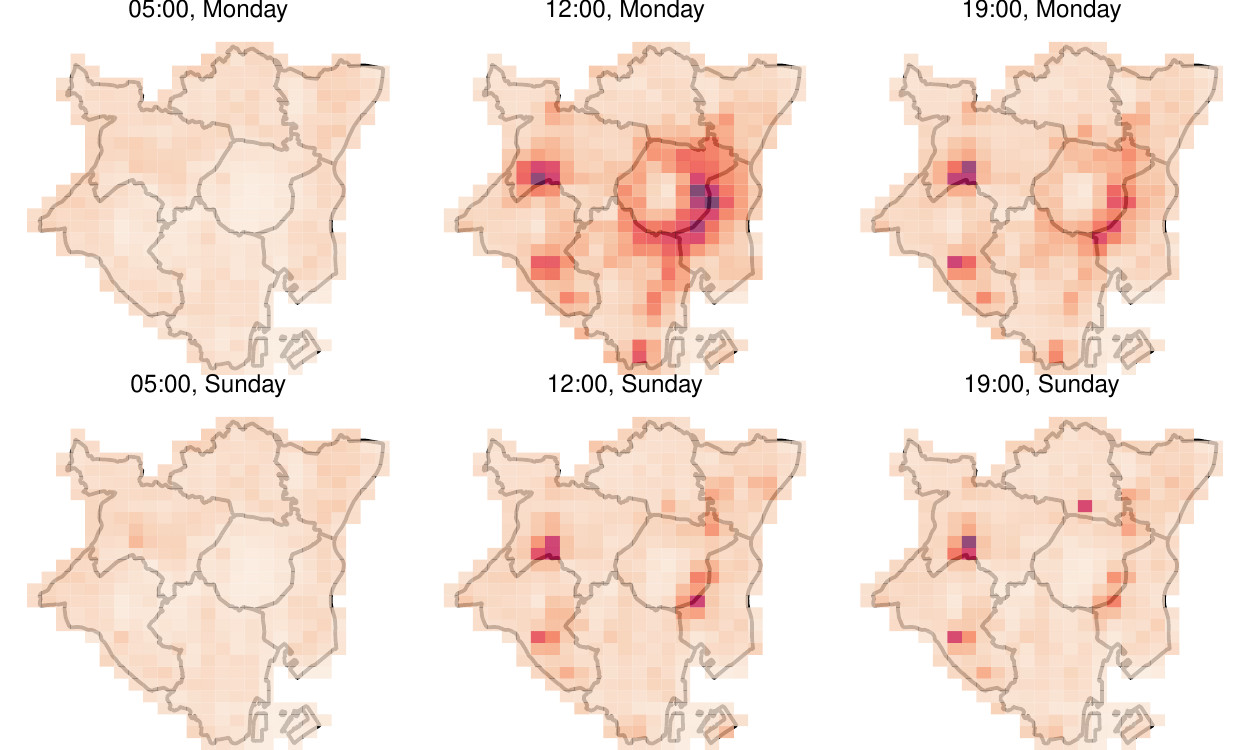}
  \end{center}
  \caption{Predicted population map by Extension II at 05:00, 12:00 and 19:00 on Monday and Sunday.}
  \label{fig:predmap}
\end{figure}

Table \ref{tab:RMSE} shows the SRMSE of the four methods for the five factors, the average RMSEs for the non-factors, and the average RMSEs and CPs for all districts. We further divided the results into errors on weekdays and holidays and overall errors. { First, the UDLM does not perform well. This suggests that the characteristics of the data are too complex to be captured by UDLM due to the spatial structure and intricate time-series patterns. For other methods, generally, the prediction accuracy and prediction interval were good, not only for the factor districts but also for the other districts, which were represented by the aggregation of factors.} This means that the factor model successfully captures the relationships among the districts.
We then examined the difference between the original and extended methods. Overall, both extensions performed better than the original. By adding the pre-holiday effect, we identified the change in the shift from weekdays to holidays. This increased the precision of the estimates for both weekday and holiday trends. Focusing on changes by region, we found large declines, particularly in Districts 2 and 5, because the extensions were able to reflect the trend that people were more likely to congregate in downtown areas on Friday nights. 
Next, we focused on how Extension I differs from Extension II. In terms of days off, Extension II outperformed by a wide margin. The tendency to stay at home instead of going out when the following day was a working day (even on holidays) was reflected in the larger changes in downtown (Districts 2 and 5) and residential areas (District 3). Figure \ref{fig:predmap} shows the population map at 5:00, 12:00 and 19:00 on the first Monday and Sunday after 152 days. This visualizes the result that the proposed method successfully captures the difference between holidays and weekdays.

\section{Conclusion}\label{sec:con}

In this study, a method for modeling and predicting spatiotemporal functional data was developed, with the primary application being population flow data. 
Our proposed method was able to use the large population data observed daily in each region for estimation and forecasting without increasing computational complexity or losing interpretability. 
First, the integration of functional data analysis and factor models (considering the time series of a small number of state variables) allowed for the computation of large data sets, and we demonstrated that the proposed method can accurately estimate and predict trends and day-off effects through
simulation studies and empirical applications.
Second, setting the factor loading matrix in the Cholesky-type matrix and considering spatial correlations and shrinkage effects in the column vectors provided a detailed and interpretable representation of the spatial structure, which is a key novelty of our work.
Furthermore, the additional incorporation of domain knowledge can improve the prediction of the proposed model. The interpretation of the corresponding terms was simple, and further information could be extracted.
Therefore, our method can facilitate the use of population data by government agencies and businesses for various applications, such as disaster prevention planning, tourism analysis, and outdoor advertising.
As more people obtain mobile devices, it will become possible to collect accurate data in finer meshes (that is, in more locations). Future studies should seek to develop approximate Bayesian methods to further reduce the computational complexity of the method.

\section*{Source Code}
The source code for the proposed methods is available at the GitHub repository (\url{https://github.com/TomWaka}).

\section*{Funding}
This study was supported in part by JSPS KAKENHI from the Japan Society for the promotion of science (grant numbers 22J21090 and 21H00699). 

\section*{Appendix: Bayesian computation procedure}\label{sec:pos}

As the prior distributions for the variance parameters, $e_s^2\sim IG(\frac{n_{e}}{2},\frac{n_{e}s_{e}}{2})$, $\lambda^2_s\sim IG(\frac{n_{\lambda}}{2},\frac{n_{\lambda}s_{\lambda}}{2})$, $\eta^2\sim IG(\frac{n_{\eta}}{2},\frac{n_{\eta}s_{\eta}}{2})$ are employed as an analytically tractable conjugate priors. 
As for the correlation kernel parameter of the Gaussian process, we set $\phi\sim IG(2,\beta)$, where $\beta = \frac{K-1}{-2\log 0.05}$ as \cite{gamerman2008spatial} did.
Concerning the day-off effect, we assume $\mu_s \sim N(0,\eta_s'R(\phi_s')).$
To ensure evolution is a stationary process, we assume $\gamma_s\sim N_{tr(-1,1)}(m_{\gamma},\sigma^2_{\gamma})$, where $G= \diag(\gamma_1,\ldots,\gamma_M)\otimes I_K$.
We set { $\psi\sim Be(\alpha_{\psi},\beta_{\psi})$} after row normalization of each $W_s$.
The data augmentation technique developed by \cite{makalic2015simple} allows the horseshoe distribution introduced in $\bm{b}_{\cdot s}$ to be re-expressed as a simple hierarchical prior. This representation also ensures that the prior conjugates and simplifies the computation of the posterior distribution. { In the experiments in this work, we set hyperparameters $n_{e}=s_{e}=n_{\lambda}= s_{\lambda}=n_{\eta}=\beta_{\psi}=\eta'=\sigma_{\gamma}=1$, $\psi=1/2$ and $m_{\gamma}=0.95$, which are standard setting in the state-space model~\cite{prado2021time}. Regarding spatial correlation, we set $\alpha_{\psi}=18$ and $\beta_{\psi}=2$, reflecting the belief that spatial correlations exist.
}

The joint posterior density of all parameters is given by\begin{align*}
   &\prod_{t,s,k} \pi \left(y_{ts}(\tau_k) \mid z_{ts}(\tau_k),e^2_s\right) 
   \prod_{t}\pi\left( \bm{z}_{t} \mid B,\bm{x}_t,\bm{\eta},\bm{\phi}\right) 
   \pi (\bm{x}_t\mid G,\bm{\mu},\bm{x}_{t-1},\Lambda) \\
   &\times \pi (B\mid \bm{\theta},\psi)
   \pi (G,\bm{e}, \bm{\theta}, \psi,\bm{\eta},\bm{\phi},\bm{\mu},\Lambda).
\end{align*}
The full conditional distributions of all parameters, except for the Gaussian process parameters, were obtained explicitly; therefore, we implemented the Metropolis algorithm within the Gibbs sampler. The concrete posterior distribution is described below.

First, the expression for $B$ introduced in (\ref{def:col}) is complicated; therefore, we reorganize it with respect to $\bm{b}_{\cdot s}$.
Define $\tilde{\bm{z}}_t^{(s)}:=(\tilde{\bm{z}}_{t,s+1}^{(s)\top},\ldots,\tilde{\bm{z}}_{tN}^{(s)\top} )^{\top}$ where $\tilde{\bm{z}}_{tj}^{(s)}:= \bm{z}_{tj} - \sum_{i\neq s}b_{ji}\bm{x}_{ti}$ for each $s\in \{1,...,M\}$, $\bm{\iota}_j \in \mathbb{R}^{N-s}$ be a vector whose $j$th element is $1$ and the others are $0$,
\begin{align*}
X_{ts} := \left(
\begin{array}{c}
\bm{\iota}_1\otimes \bm{x}_{ts} \\
\bm{\iota}_2\otimes \bm{x}_{ts} \\
\vdots \\
\bm{\iota}_{N-s}\otimes \bm{x}_{ts}
\end{array}
\right).
\end{align*}
Then, we obtain $\tilde{\bm{z}}_t^{(s)} = X_{ts}\bm{b}_{\cdot s}+\bm{\nu}_t$.
Once the standard form of the regression using $\bm{b}_{\cdot s}$ is obtained, the remainder of the calculation is simple. Here, we present the full conditional distribution.

\begin{itemize}   
   \item[-] (Sampling from $\bm{z}_t$) \ \ The full conditional distribution of $\bm{z}_t$ is $N(\bm{m}_{z_t}, \Sigma_{z_t})$, where
    \begin{align*}
    \bm{m}_{z_t}       & = \Sigma_{z_t}\left(E^{-1}\bm{y}_{t}+ \mathrm{blockdiag} (\tilde{R}_1^{-1},...,\tilde{R}_N^{-1} )(B\otimes I_K) \bm{x}_t \right),  \ \ \   \\
    \Sigma_{z_t}  & = \left( E^{-1} + \mathrm{blockdiag} (\tilde{R}_1^{-1},...,\tilde{R}_N^{-1} ) \right)^{-1}, \ \ \  \\
    E             & = \diag(e_s^2)\otimes I_K.
    \end{align*}

   \item[-] (Sampling from $\bm{x}_t$) \ \ The full conditional distribution of $\bm{x}_t$ is $N(\bm{m}_{x_t}, \Sigma_{x_t})$, where
    \begin{align*}
    \Sigma_{x_t}  & = \left( (B\otimes I_K)^{\top}\mathrm{blockdiag} (\tilde{R}_1^{-1},...,\tilde{R}_N^{-1} )(B\otimes I_K)+\Lambda^{-1} + G^{\top} \Lambda^{-1} G  \right)^{-1}, \\
    \bm{m}_{x_t}& = \Sigma_{x_t}\Big((B\otimes I_K)^{\top}\mathrm{blockdiag} (\tilde{R}_1^{-1},...,\tilde{R}_N^{-1} )\bm{z}_t+ \Lambda^{-1}G\bm{x}_{t-1}\\
    &\ \ \ \ \ \ \ \ \ \ \ \ \ + G \Lambda^{-1} \bm{x}_{t+1} +D_t\Lambda^{-1}\bm{\mu} - D_{t+1} G^{\top}\Lambda^{-1}\bm{\mu} \Big).
    \end{align*}

   \item[-] (Sampling from $G$) \ \  The full conditional distribution of $\gamma_s$ is $N_{tr(-1,1)}(\tilde{m}_{\gamma},\tilde{\sigma}^2_{\gamma})$, where
    \begin{align*}
    \tilde{m}_{\gamma}         & = \tilde{\sigma}^2_{\gamma} \left(\sum_{t=2}^T \frac{ ( \bm{x}_{ts}- D_t\bm{\mu}_s )^{\top}\bm{x}_{t-1,s}}{\lambda^2_s} + \frac{m_{\gamma}}{\sigma^2_{\gamma}}\right),    \\
    \tilde{\sigma}^2_{\gamma}  & = \left(\sum_{t=2}^T  \frac{\bm{x}_{t-1,s}^{\top}\bm{x}_{t-1,s}}{\lambda^2_s} + \frac{1}{\sigma^2_{\gamma}}   \right)^{-1}  .
    \end{align*}

   \item[-] (Sampling from $e_s^2$) \ \ The full conditional distribution of $e^2_s$ is\begin{align*}
       IG\left(\frac{n_e+TK}{2},\frac{n_{e}s_{e}+\sum_{t=1}^T\| \bm{y}_{ts}-\bm{z}_{ts}\|_2^2 }{2}\right).
   \end{align*}

    \item[-] (Sampling from $\lambda^2_s$) \ \ The full conditional distribution of ${\lambda}^2_s$ is\begin{align*}
       IG\left(\frac{n_{\lambda}+(T-1)K}{2},\frac{n_{\lambda}s_{\lambda}+\sum_{t=2}^T\|\bm{x}_{ts}-\gamma_s\bm{x}_{t-1,s}- D_t\bm{\mu}_s\|_2^2 }{2}\right).
   \end{align*}

    \item[-] (Sampling from $\eta_s^2$) \ \ The full conditional distribution of $\eta_s^2$ is\begin{align*}
       IG\left(\frac{n_{\eta}+TK}{2},\frac{n_{\eta}s_{\eta}+\eta_s^2\sum_{t=1}^T(\bm{z}_{ts}-(\bm{b}_{s\cdot}\otimes I_K ) \bm{x}_t)^{\top}\tilde{R}^{-1}_{\phi_s}(\bm{z}_{ts}-(\bm{b}_{s\cdot}\otimes I_K ) \bm{x}_t) }{2}\right).
   \end{align*}

   \item[-] (Sampling from $\phi_s$) \ \ The full conditional distribution of $\phi_s$ is not written in analytic form. Hence we sample $\phi_s$ using the random-walk Metropolis-Hastings method with acceptance rate \begin{align*}
        \min \left[1, \frac{ \tilde{\phi}_s^{-3}\exp\left(-\frac{\beta}{\tilde{\phi}_s}\right)\prod_{s}\det(\tilde{R}_{\tilde{\phi}_s})^{-\frac12}\exp \left\{ -\frac12 \left(\bm{z}_{ts} - (\bm{b}_{s\cdot}\otimes I_K )\bm{x}_t\right)^{\top} \tilde{R}_{\tilde{\phi}_s}^{-1}\left(\bm{z}_{ts} - (\bm{b}_{s\cdot}\otimes I_K )\bm{x}_t\right)  \right\}}{\phi_s^{-3}\exp\left(-\frac{\beta}{\phi_s}\right)\prod_{s}\det(\tilde{R}_{\phi_s})^{-\frac12}\exp \left\{ -\frac12 \left(\bm{z}_{ts} - (\bm{b}_{s\cdot}\otimes I_K )\bm{x}_t\right)^{\top} \tilde{R}_{\phi_s}^{-1}\left(\bm{z}_{ts} - (\bm{b}_{s\cdot}\otimes I_K )\bm{x}_t\right)  \right\}}\right].
   \end{align*}

   \item[-] (Sampling from $\bm{\mu}_s$) \ \  The full conditional distribution of $\bm{\mu}_s$ is $N(\bm{m}_{\mu_s},\Sigma_{\mu_s})$, where
    \begin{align*}
    \bm{m}_{\mu_s} & = \Sigma_{\mu_s} \lambda_s^{-2} \sum_{t=2}^TD_t(\bm{x}_{t}- G\bm{x}_{t-1} ) ,\\
    \Sigma_{\mu_s} & =  \left( \sum_{t=2}^TD_t^2 \lambda_s^{-1}\right)^{-1}.
    \end{align*}

   \item[-] (Sampling from $B$)
   The full conditional distribution of $\bm{b}_{\cdot s}$ is $N(\bm{m}_b, \Sigma_b),$ where \begin{align*}
       \bm{m}_b  &= \Sigma_b\sum_{t=1}^T X_{ts}^{\top} \mathrm{blockdiag}(\tilde{R}_1^{-1},...,\tilde{R}_N^{-1})  \tilde{\bm{z}_t}^{(s)}\\
       \Sigma_b  &= \left(\upsilon^2\bm{\theta}_s^{-2}Q_s + \sum_{t=1}^T X_{ts}^{\top} \mathrm{blockdiag}(\tilde{R}_1^{-1},...,\tilde{R}_N^{-1})  X_{ts} \right)^{-1}.
   \end{align*}

    \item[-] (Sampling from $\theta_s^2$) \ \ The full conditional distribution of ${\theta}_s^2$ is\begin{align*}
       IG\left(\frac{N-s+1}{2},\frac{\bm{b}_{\cdot s}^{\top}Q_s(\psi)\bm{b}_{\cdot s} }{2\upsilon^2}+\frac{1}{\zeta_s}\right).
   \end{align*}
   
   \item[-] (Sampling from $\zeta_s$) \ \ The full conditional distribution of ${\zeta}_s$ is\begin{align*}
       IG\left(1,\frac{1}{\theta_s^2}+1\right).
   \end{align*}
   
    \item[-] (Sampling from $\upsilon^2$) \ \ The full conditional distribution of $\upsilon^2$ is\begin{align*}
       IG\left(\frac{1+\sum_{s=1}^M N-s}{2},\sum_{s=1}^M\frac{\bm{b}_{\cdot s}^{\top}Q_s(\psi)\bm{b}_{\cdot s} }{2\theta_s^2}+\frac{1}{\nu}\right).
   \end{align*}
   
   \item[-] (Sampling from $\nu$) \ \ The full conditional distribution of $\nu$ is\begin{align*}
       IG\left(1,\frac{1}{\upsilon^2}+1\right).
   \end{align*}
   
   { 
   \item[-] (Sampling from $\psi$) \ \ The full conditional distribution of $\psi$ is not analytically available. Hence we implement random-walk Metropolis-Hastings with acceptance rate \begin{align*}
       \min \left[1, \frac{\tilde{\psi}^{\alpha_{\psi}-1}(1-\tilde{\psi})^{\beta_{\psi}-1}\prod_{s=1}^M\det\left(Q_s^{-1}(\tilde{\psi})\right)^{-\frac12}\exp\left( -\frac{\bm{b}_{\cdot s}^{\top}Q_s(\tilde{\psi})\bm{b}_{\cdot s}}{2\upsilon^2\theta_s^2}\right)}{\psi^{\alpha_{\psi}-1}(1-\psi)^{\beta_{\psi}-1}\prod_{s=1}^M\det\left(Q_s^{-1}(\psi)\right)^{-\frac12}\exp\left( -\frac{\bm{b}_{\cdot s}^{\top}Q_s(\psi)\bm{b}_{\cdot s}}{2\upsilon^2\theta_s^2}\right)} \right].
   \end{align*}
    } 

\end{itemize}

\bibliographystyle{chicago}
\bibliography{reference}

\end{document}